# A Blockchain and Artificial Intelligence based System for Halal Food Traceability

ABDULLA ALOURANI[1] & SHAHNAWAZ KHAN[2]

[1]Department of Computer Science and Information,
College of Science in Zulfi, Majmaah University
Al-Majmaah, 11952, Saudi Arabia
[2]Faculty of Engineering, Design and Information & Communications Technology
Bahrain Polytechnic, Bahrain
E-mail: [1]a.alourani@mu.edu.sa, [2]shahnawaz.khan@polytechnic.bh

**Abstract:**

The demand of the halal food products is increasing rapidly around the world. The consumption of halal food product is just not among the Muslims but also among non-Muslims, due to the purity of the halal food products. However, there are several challenges that are faced by the halal food consumers. The challenges raise a doubt among the halal food consumers about the authenticity of the product being halal. Therefore, a solution that can address these issues and can establish trust between consumers and producers. Blockchain technology can provide a distributed ledger of an immutable record of the information. Artificial intelligence supports developing a solution for pattern identification. The proposed research utilizes blockchain an artificial intelligence-based system for developing a system that ensure the authenticity of the halal food products by providing the traceability related to all the operations and processes of the supply chain and sourcing the raw material. The proposed system has been tested with a local supermarket. The results and tests of the developed solution seemed effective and the testers expressed interest in real-world implementation of the proposed system.

**Introduction**

The demand of the halal food and concerns regarding the traceability of halal food are increasing worldwide (Tan et al., 2022). The term 'halal' is an Arabic language word. It translates to 'permissible' or 'lawful' in English language. The consumers of the halal food and halal products are Muslims primarily. However, as because halal foods and products substantiate high standard of cleanliness and purity throughout the life cycle from the origin of the constituents to the consumption by the end user. Therefore, the acceptance of the halal foods and products are increasing in the non-muslim community as well (Rohmah et al., 2019). Muslim consumers do not consume any items that is not halal according to Islamic principles. A product needs to satisfy certain criteria to be halal according to Islamic principles. These criteria do not only include the ingredient and origin of the constituents but also include the process, and supply chain operations, etc. (Khan, 2024; Riaz & Chaudry, 2018). For example, a cow is a halal animal but if the method of slaughter is not Islamic then its meat after slaughter will not be halal. Similarly, if the halal food products are mixed with non-halal products during the production or transportation, they may contaminate halal products and therefore, halal product will no longer be considered as halal. Therefore, it becomes of utmost important that standard Islamic principles have been followed throughout the life cycle of the halal food or product such as raw material sourcing, processing, production, logistics and transportation, etc. (Al-Teinaz et al., 2020; Riaz & Chaudry, 2018).

As per the studies (Statista, 2024), the total consumer spending on all halal products surpasses two trillion U.S. dollars on products and services within the food, fashion, travel, pharmaceutical, cosmetic, and media/recreation industries. This market is expected to expand to approximately 2.8 trillion dollars by the year 2025. Moreover, global expenditure by Muslims on halal food reached approximately 1.26 trillion U.S. dollars in 2021. The projections indicates that global Muslim spending on halal products will increase by 7.1% annually from 2024 to 2025 and the halal food reached expenditure figure will rise to 1.67 trillion U.S. dollars by 2025 (Statista, 2024). However, the halal food consumers are primarily concerned about the authenticity of the product being halal. Since, there are several challenges in the current supply chain that limits traceability information while tracing the halal food supply chain

(Tan et al., 2022). Therefore, ensuring the authenticity and compliance of halal certification across the entire supply chain has posed numerous challenges, including issues related to transparency, traceability, and trust.

There is no standard system available to provide such facilities that guarantees the authenticity and compliance with the Islamic principles. Therefore, it has become increasingly important to develop such a system that can establish the trust between the consumers and the producers of the halal foods or products. There are several factors that pose challenges during the halal food production and supply chain. These factors include lack of a universally accepted halal certification, and variability of the halal certification standard due to regional differences and differences of opinions among Islamic schools of thought. Therefore, it has become more and more complex to ensure the authenticity and compliance of halal certifications.

To address these challenges and issues, a system is needed that can provide better traceability of the information related to the production process, sourcing of the constituents and logistics operations as well as storage and supervision during the supply chain. This research proposes a system using blockchain and artificial intelligence technologies. The blockchain (K. Shahnawaz et al., 2022; Sharma et al., 2022) technology provides direct access to verify the public blockchain ledger. The decentralize nature of the blockchain network and its ability to provide immutable record of ledger make it an ideal solution for a transparent and reliable ledger with the feature of traceability to verify the authenticity and provenance of goods. Therefore, with this level of transparency, the proposed system will aid in building the trust between consumers and producers. Artificial intelligence has the ability to process vast amount of data and identify patterns. Although, artificial intelligence has been applied in various fields such as machine translations (Shahnawaz & Mishra, 2015; S. Shahnawaz & Mishra, 2012), image processing (Bashir et al., 2017), transactional fraud detection (Khan et al., 2022), educational system (Khan, Al-Dmour, et al., 2021) and organizational performance (Khan et al., 2020) transformations, etc. However, this research will be utilizing AI for the purpose of pattern identification. The proposed system will be using quick response (QR) codes for tracking and tracing the information related to the halal food.

This paper has been organized into six sections. The next section reviews the literature and collects the information on how the halal food traceability systems have been implemented or what limitations they have (if any). Literature review section also identifies the criteria utilized to evaluate the supply chain traceability systems. These criteria have been synthesized and have been utilized for system evaluation. The following blockchain traceability section provides brief details on some of the differences among Islamic schools of thoughts for halal food standards. The research methodology section explains the system design and discusses the system implementation. The next section is results and discussion section that discusses the results. Section six is the conclusion and future work section which provides the conclusion and insights into the future work.

**Literature Review**
Blockchain technology has been recognized as a powerful tool for creating decentralized and tamper-proof ledgers within supply chains. The decentralized nature of blockchain technology ensures transparency and prevents data tampering, thus ensuring the integrity of the halal food supply chain. Several studies have explored its application in the halal food industry. The application of blockchain has been observed in several areas such as banking, medical services, supply chain etc. A research study (Ahamed et al., 2024) suggest that by integrating blockchain with special tags such QR codes or radio frequency identification (RFID) can ensure the consumers to trust the authenticity of the halal food products. The study demonstrates that the use of special tags, along with smart contracts has given the accessibility of the information to the end consumers. But as the smart contracts have a predefined specifications and data, the issue of universally accepted halal certification persists. Another similar empirical research, conducted as a case study on a halal chicken meat supplier, also suggest that QR code and blockchain based supply chain system can help in developing a transparent and trustworthy system for the halal food consumers (Susanty et al., 2024). However, the proposed system (Susanty et al., 2024) has certain limitation such as the consensus process is manual, and geographic location of the resourcing and productions have not been considered. Moreover, the research focuses on the guiding the policymakers for resolving the issues related to the halalness of the product. The fraudulent activities and halal standard compromising activities during the supply chain remains a major concern among the consumers.

The consumers of the halal food and other halal products seek to trust the government in providing strict regulations and providing them a solution that can help them in the traceability of the halal food and products (Karyani et al., 2024). However sometimes, the quality of the halal product has been compromised by some bad actors involved in the halal food supply chain. A research study (Mahsun et al., 2023) analyzes the widespread counterfeiting of the halal products and develops the concept of a blockchain based supply chain to address the counterfeiting issue. A research study (Wahyuni et al., 2024) identifies different risks and points of failure to comply with the halal food certifications in beef-meat supply chain. This research proposes a blockchain framework to minimize the risk and provide more reliability on beef-production with regard to halal standard compliance during the production process. A research study (Rahman, 2024) addresses the regulatory framework for integrating blockchain and artificial intelligence. An analysis (Ellahi et al., 2023) of various research study exploring the application of blockchain and other technologies illustrates that the food supply chain can be transformed to be more transparent and traceable by applying web 3.0 technologies such as blockchain. A qualitative research study (Dashti et al., 2024a) interviews the policy makers and the halal food producers for understanding their perspective with regarding to transforming the food supply chain for providing better traceability of the information for the purpose of creating transparency and trustworthiness about the authenticity of the halal food between the consumers and the producers. The study identifies that the processes related to the farms or cultivation, suppliers, processors are critical components. The information related to these can be made available to the end consumers via traceability and the application of innovative technologies such as blockchain.

A research study (Alamsyah et al., 2022) proposed a blockchain-based system that enabled stakeholders to record and verify halal certificates and transactions securely. The application of artificial intelligence has also become regular in supply chain processes. For example, a research study (Abidin & Perdana, 2020) has proposed a system that utilized AI-based image recognition to track and authenticate halal meat products at each stage of the supply chain. The proposed research also utilizes blockchain for providing secure and transparent record-keeping. Machine learning is a specialized area of artificial intelligence that can make a machine intelligent enough to automatically classify the constituents of a product (Khan, Thirunavukkarasu, et al., 2021; Thirunavukkarasu et al., 2021) into halal and non-halal constituents. A research (Dashti et al., 2024b) proposes a machine learning based model for automating the halal certification based on the constituents of the product. However, as such a system may be quite useful but it has certain limitation and can not be applied practically because of the issues that may arise during the supply chain and can contaminate the halal food product as discussed earlier. A research study (Chandra et al., 2019) proposes a framework that utilized blockchain, smart contracts, and IoT devices to ensure the integrity of the halal supply chain to counter the frauds related to halal food supply chain. A comprehensive review by (Bux et al., 2022) of existing literature on blockchain technology in the food supply chain highlights the need for further research to develop practical solutions that address the unique requirements of the halal industry. A research study (Tieman et al., 2019) advocates that a system can be developed by leveraging AI and blockchain technology which can enhance consumer trust, mitigate fraud, and promote compliance with Islamic dietary guidelines.

Therefore, it can be conclude based on the above discussion that the halal food supply chain requires an innovative solution that provide transparency, and traceability of the information related to the producers, logistics and suppliers. The consumers are striving to find a trustworthy solution. The discussion also indicates that blockchain and artificial intelligence have the potential to provide such solution. These technologies can help in developing a supply chain solution that can provide transparency, security, and information verification. The process related to monitoring and verification can be automated. Blockchain technology can help in developing a tamper-proof documentation for traceability and mitigating the fraud. The following section provides more details on blockchain traceability.

**Blockchain traceability**

Halal food certification is a key factor for Muslim consumers while purchasing raw food items and off-the-self food items. Halal food standard might vary slightly depending on the Islamic school of thought. However, the basic principles remain the same. In general, halal food is free from pork and its derivatives, slaughtered in Islamic way, free from intoxicants, haram ingredients, and unethical practices. It is just certain items are permitted within a school of thought while other consider it impermissible. However, these items do not contaminate the other food. While any cross-contamination with universally agreed upon impermissible items contaminate the whole food (Kurniawati & Cakravastia, 2023). These differences primarily arose based on the different interpretations by the scholars of different Islamic schools of thought (Hanafi, Maliki, Shafi'i, and Hanbali). For example, Hanafi and Maliki permits the consumption of all fish with scales while Shafi'i, and Hanbali require that fish have both scales and fins. Therefore, to implement the right levels of transparency by providing traceability information, the trust will increase between the consumers and the producers.

Ensuring that the halal food products are actually authentic, and there have not been any issue that can tamper with the authenticity of the halal food products, is a challenging task. Halal certificates play a crucial role in ensuring the authenticity of the halal product. However, one of the major challenges for the producers and consumers is accepting a halal certificate as the trusted one (Abidin & Perdana, 2020; Alamsyah et al., 2022). There are several halal certificates based on the areas for example, halal food standard developed by Standards and Metrology Institute for Islamic Countries, Indonesian Ulema Council (MUI) certificate in Indonesia, JAKIM certificate in Malaysia, halal certificates accredited by GCC, halal certificates in Europe and in North America, etc. Though, as per Islamic dietary principles and regulations, the requirements for the food being halal are same regardless of the location. However, due to the requirements and compliance with the local authorities, halal certificate standards might not be exactly same in two different places (Abidin & Perdana, 2020). For example, it is highly likely that halal certificate issued by MUI Indonesia and the one issued in Europe might have slightly different standards. All of these ambiguities can lead to the inconsistencies in the process and standards of the halal certificate. Therefore, a universally accepted halal certificate can be a solution to these issues.

Halal food traceability information becomes a key factor for Muslim consumers. The proposed system provides the information related to the source, components, production and distribution, etc. Traceability becomes important because it also provide information related to the compliance with the halal food standards. Moreover, government regulations around the world enforces the organizations to maintain food standards. Therefore, traceability in halal food industry has become an obligation to protect the chastity and halal food standard throughout supply chains. It has become essential for enhancing the efficiency and visibility of the halal food standards among the consumers and the organizations that brings the food to them. Halal food consumers require certain trust and transparency starting from the raw resource, compositions, production and distribution. The implementation of the blockchain in the proposed system would be able to provide perfect conditions, for the organization and consumers, to implement the right levels of transparency by providing traceability information.

Therefore, given the aforementioned considerations, it is imperative to implement robust traceability measures to ensure the authentic halal status of products. The proposed system is capable of establishing the authenticity, constituents and geographic journey of the product throughout its supply chain. It also fulfills the need of the government regulations that have enforced the industries to implement a traceability system in the food industry. The proposed system provides the end-consumer with the enough amount of details on the product development, constituents, processing, production, storage and distribution supply chains. This level of traceability develops a solid trust and transparency about the halal authenticity of the product.

As the halal-food consumers demands assurance that the product, marketed as halal, has complied with the halal food standards. In the proposed system, although, there are no provisions to provide information about a specific Islamic school of thought but, the details about the constituents, storage and sources required for production of the food-item can be traced. Therefore, there is no requirement to put a specific label as per the Islamic schools of thought. The halal-food consumer can make decision based on the information provided through traceability.

**Research Methodology**

In developing the proposed system, the primary problem is universally accepted halal certificate. There are two main solutions to overcome the issue of the universally accepted halal certificate. First solution is to rely on the standards of the existing certificates, but keeping track of the different activities and standards at different stages of the halal food production and ensuring the same through the distribution channels, logistics and retails. Second solution is to develop a universally accepted global halal certificate. Though first solution does not look extremely authentic (which is very difficult to achieve because of the certain differences among scholars) but is feasible (during the life-span of this project) to implement. Second solution makes the process exceptionally authentic but hardly feasible to implement during the life of the project. Therefore, to resolve this dilemma, this research has implemented an alternate solution which is to provide enough details to the end-consumers along with the information about halal certification body at different stage of the supply chain. During the supply chain, it is possible that the food has been produced based on the standards set by one halal certification body and transported and stored by the organization that uses different halal food standard.

The proposed project has been divided into several steps such as research problem formulation, data collection and analysis, developing and evaluating the system. Research project has been initiated by identifying and articulating the specific challenges and gaps pertaining to transparency and integrity in the halal food supply chain. Then precise research questions have been developed that align with the objectives of this research. A comprehensive literature review has been conducted to obtain an in-depth understanding of the current research and knowledge on AI, blockchain, and their applications in supply chain management, particularly in the context of the food industry. The next step determines the specific data required to address the research aims, such as certification records, transaction data, and product information. A model that integrates AI and blockchain technologies to facilitate transparent tracking of the halal food supply chain will be developed. AI algorithms for monitoring and verification via QR (quick response) codes, and leverage blockchain for decentralized and tamper-proof documentation will be implemented.

The research methodology followed in this research project has been divided into four primary phases. These phases are business process understanding, business transformation, system implementation and evaluation. The first phase is understanding the business process and identifying the key points for information traceability, second phase is transforming the business process into blockchain based system, third phase is implementing the proposed blockchain based system and the fourth phase is evaluating the proposed system. The following figure (see figure 1) illustrate the high-level system diagram of the proposed system.

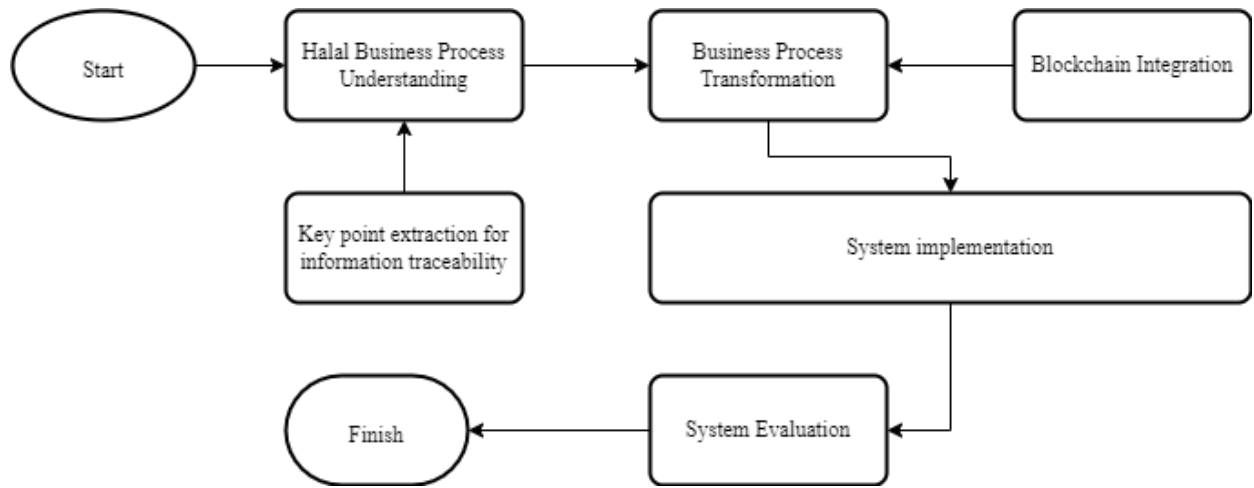

*Figure 1: Proposed Halal Food Traceability System*

The first phase of the proposed system is to understand the business process around halal food industry and then identifying the key points for information collection which can be used for providing traceability information. The supply chain of halal food products is a complex process. It starts with ensuring that the raw material used for the production of the halal food is halal. It can involve verifying the ingredients, origin of animals, methods of slaughter and handling of all these. This step is crucial in certifying the product halal and also for data collection to maintain the transparency and traceability.

**System Design and Implementation**

There are several stakeholders involved in the halal product supply chain. For developing a trustworthy system, the information contributions from all stakeholder are essential. The proposed system has three primary functions related to the information form the halal food production processes. The first function is data collection about the processes, second is confirming the transactions or processes and third is tracing the information. To make the system transparent, the system is able to provide a track of the traceability information from the origin and journey of the raw materials throughout the halal food supply chain. Halal food supply chain includes a variety of stakeholders. These stakeholders include farmers, producers, retailers, and consumers etc. There are four primary stakeholders in the halal food supply chain which are cultivators, makers, merchants and consumers. The use-case diagram (see figure 2) demonstrates the high-level relationships between the stakeholders, and the system. It illustrates that majority of the stakeholders (cultivators, makers, and merchants) have the responsibility for collecting authentic information related to the different processes, recording that information into the blockchain system, and confirming that the information providing is authentic. Consumers along with other stakeholders have the privileges to trace the information to determine the halal authenticity of the product.

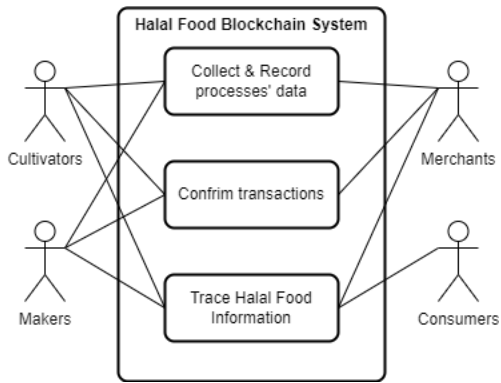

*Figure 2: Use-case diagram demonstrating the relationships between the stakeholders, and the system*

First category of the stakeholders includes farmers and producers who provide halal raw materials such as poultry, livestock, and agricultural produce, let us call them cultivators. When it comes to raising halal animals and growing crops, Islamic principles must be followed while feeding, slaughtering and handling. The responsibilities of the cultivators include adhering to the halal food practices in animal husbandry, slaughter, and crop cultivation. The second category includes processors, producers, packagers, and manufactures, let us call them makers. These entities process and package the halal food. Their responsibilities include adhering to the halal food practices during the production process. For example, making sure if the halal ingredients and proper equipment have been used. Similarly, the packaging must be done in a way that the product does not contaminate and it clearly indicates the halal status. The third category includes the stakeholders who are responsible for bringing the food to the consumers. These includes distributors and retailers, let us call them merchants. The merchants must make sure that the proper halal conditions have been adhered during the transportation and the storage of the halal products, hence, maintaining the integrity of the product throughout the life supply chain. The fourth category of the stakeholders is consumers. However, one more important stakeholder is halal food certification body which has been discussed in previous sections (blockchain traceability and research methodology). There are several other secondary stakeholders which may play a crucial role in this regard. These include government and regulatory bodies, non-governmental organizations, financial institutions, etc. However, because of the scope limitation, those have not been included in the further discussion.

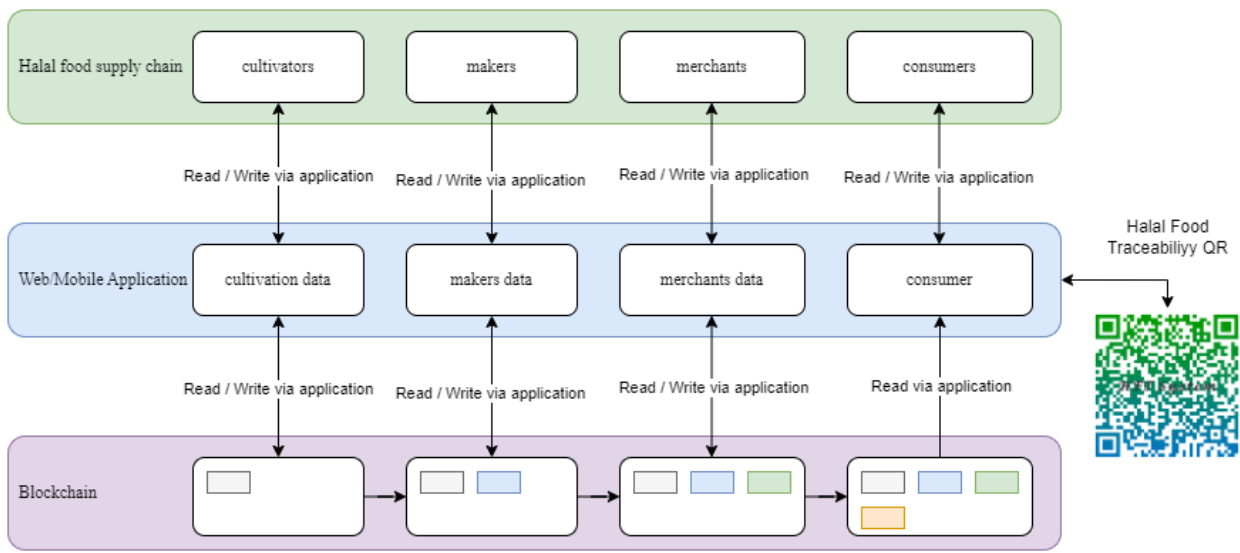

*Figure 3: Reading/Writing data from/to Blockchain*

The proposed halal food traceability system consists of three interconnected levels for collaboration among all stakeholders (see figure 3). Each stakeholder interacts with the system through an application (web or mobile) interface. The above diagram demonstrates how the halal food information which can be written, read and traced using QR code. The application provides the capability to write the information to and to read the information from the blockchain infrastructure. The following steps explains the operational details of the proposed system:

- The cultivators record the animal husbandry information, crop production information, processing and handling information, halal certification information and traceability records.
- The makers record the ingredient information, production process information, employee training records related to halal food management, halal certification information and traceability records.
- The merchants record the product information, purchase and receipt information, storage and handling information, distribution and sale information, and halal certificate compliance verification information.
- The consumer can verify the product for its halal authenticity using the application.

In the above process of recording various kinds of information, one can argue that in the above steps, there is no information on the role of governmental or national agencies etc. However, those are important from legal point of view but does not contaminate the traceability records.

The following swim lane diagram (see figure 4) demonstrates how and what data is being recorded from different activities. System records name, location, halal certification number, facility manager's contact information from the cultivators related to the facility information. System records specific type of halal raw material (e.g., poultry, livestock, agricultural produce), animal husbandry practices (such as feed), slaughtering methods if applicable, harvesting and processing process, and traceability record. Makers records the product information such as ingredients, production process, halal certification number, packaging, cultivators' traceability record, own traceability record, production date, batch, and quality control records including employees' halal food management training. The merchants record purchased and receipt information such as purchase date, invoice number, supplier contact information, storage conditions, locations, and handling procedures, halal certification number, makers' traceability record. At this stage all the required information, for providing halal food traceability information, has been recorded. Hence, a QR code is created and associated with the halal food products for which the information has been collected.

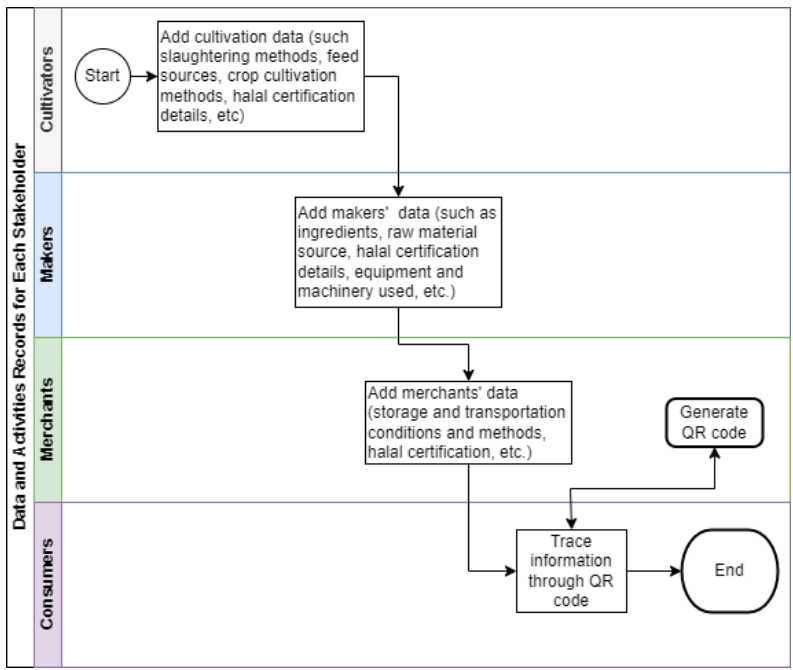

*Figure 4: Swimlane diagram for Data and Activities Records for Each Stakeholder*

Integrating blockchain technology with a complex system such as supply chain requires a deep understanding of its underlying principles. To make the integration successful, the level of abstraction, functional features and implementation details must be considered in advance. The proposed system is an end-to-end solution. While the end users are not technical. Therefore customization, compatibility and integration (with other technologies being used such as web framework in this case) of the blockchain infrastructure is crucial. The integrated framework should be able to invoke smart contracts when required. Therefore, this research develops its own blockchain infrastructure and utilizes Proof of Stake (PoS) consensus mechanism. Although, there many existing blockchain platform that provides similar functionalities, however due to their limitation such as limited flexibility in terms of adding or modifying features, limited control over data privacy, etc. a custom blockchain solution can be highly tailored, scalable and cost-effective.

The prototype of the proposed system has been implemented using Python programming and Flask framework. The prototype blockchain has also been designed using Python. The system defines a data structure called 'Block'. It specifies the structure of the block include the data being recorded, timestamp, previous hash and the calculated hash. System contains a primary Python class for blockchain structure and calls it Blockchain. This class manages the chain of blocks, and provides provisions to create genesis block, getting the latest block and adding new blocks. However, direct interaction with the blockchain requires some technical skills which the system considers negatively available in the stakeholders. Therefore, a web interface has been developed to easily interact with the proposed system. This interface can be further improved and a web or mobile application can be developed. The system utilizes the flask framework for smooth interaction with the blockchain. Python based artificial intelligence (AI) library (pyzbar) has been utilized to detect QR (quick response) codes and QR code generator library (qrcode) has been used for generating QE codes. This research utilizes QR code for providing traceability information. AI-based algorithm (utilizing python libraries) has been designed and implemented to automate monitoring and verification processes of halal certification using QR code detection, information extraction and generation (Alamri & Khan, 2023). It improves the efficiency and accuracy in providing the traceability information of the halal food products. Blockchain technology has been utilized to create a decentralized and tamper-proof ledger, ensuring transparency and integrity of transactions within the halal food supply chain.

**Results and Discussion**

The information of a local poultry farm which directly supplies the poultry to the supermarket was collected for testing the prototype. All the information being recorded is textual information. Location of the facility has been recorded using the latitude and magnitude. This information can be directly fed to the map application to get the actual location of the facility. As soon as the information is added to the system, a unique traceability number for cultivators is generated and recorded in the system. This number is associated the specific batch or lot number. Similarly, the maker records the information related the manufacturing and processing information and their traceability number associated the specific batch or lot number. The next stakeholder who is responsible for adding the halal traceability information is merchants. The merchants record the storage, transportation and distribution information. Once all the information has been collected and recorded, the next phase is with the consumers to access the recorded information.

The halal food traceability system has been evaluated on several criteria. The list of the requirements has been developed based on the blockchain evaluation information collected during the literature review and based on the technical aspects of the proposed systems. The following table (see table 1) illustrates the system performance evaluation functional and non-functional requirements, tests conducted and their results. The prototype of the proposed system has been tested with multiple end consumers. The selection of the tester at end-consumer is primarily based on the expertise in technical skills. If the user has high technical skills, then the candidate will not be selected if the candidate is very low skilled at technical level. The prototype was tested by multiple non-technical skills. All the testers have given positive feedback and eager to try the future versions of the prototype. The following tables 1 (see table 1) list the different types of functional and non-functional requirements and the measurement criteria for the requirement to be successfully completed. The table 2 (see table 2) provides a list of the tests conducted and test results for each of the functional and non-functional requirements. For testing the security requirement, automated tools (Nmap and Nessus) have been used.

*Table 1: Functional and non-functional requirements and the measurement criteria*

| Criteria | Description | Measurement Criteria |
|---|---|---|
| **Functional Requirements** | | |
| Data Integrity | Ensuring the immutability of data records. | Compare recorded data with final traceability results. |
| Transparency | Provide an open and auditable process for supply chain traceability. | Review the transparency of the supply chain stream. |
| User Access Control | Implement a secure and flexible user access control mechanism. | Verify that users have appropriate access to data based on their roles. |
| Interoperability | Ensure the system is accessible from various devices. | Test whether users can read and write to the system using different devices. |
| Scalability | The system must be capable of handling a large volume of transactions. | Measure the maximum number of concurrent transactions the system can handle. |
| Cost-Effectiveness | Ensure that operational costs, such as transaction fees, are reasonable. | Calculate the average cost per transaction. |
| **Non-Functional Requirements** | | |
| Performance | The system should respond to user requests in a timely manner. | Measure response times for various user actions. |
| Security | The system must protect sensitive data from unauthorized access. | Conduct vulnerability assessments and penetration testing. |
| Reliability | The system should be highly available and have minimal downtime. | Track system uptime and mean time between failures. |
| Usability | The system should be easy for users to understand and use. | Conduct user interface evaluations and usability testing. |

*Table 2: Tests and test results for functional and non-functional requirements*

| Requirement | Test | Test Results |
|---|---|---|
| **Functional Requirements** | | |
| Data Integrity | Compare recorded data with final traceability results. | No discrepancies found between recorded and final data. |
| Transparency | Review the supply chain stream for transparency. | All relevant information about suppliers, processes, and certifications is accessible. |
| User Access Control | Attempt to access data with unauthorized credentials. | Access denied for unauthorized users. |
| Interoperability | Test system access using different devices and browsers. | System functions correctly on all supported devices and browsers. |
| Scalability | Simulate a high volume of transactions and measure system performance. | System handles the simulated load without significant performance degradation. |
| Cost-Effectiveness | Calculate the average cost per transaction. | Transaction costs are within acceptable budget constraints. |
| **Non-Functional Requirements** | | |
| Performance | Measure response times for various user actions. | Response times are within acceptable thresholds. |
| Security | Conduct vulnerability assessments and penetration testing. | No critical vulnerabilities identified using Nmap and Nessus |
| Reliability | Monitor system uptime and mean time between failures. | System uptime exceeds 99.9%. |

| Usability | Conduct user interface evaluations and usability testing. | The users complete tasks efficiently and with minimal errors or no errors. |

**Conclusion and Future works**

This research has successfully designed and implemented a halal food traceability system using blockchain and artificial intelligence. The proposed designed is a general model. The proposed model can be molded to fit a specific kind of food category (or something else) where traceability can be easily seen. The proposed system has been implemented using Python and flask framework and QR code AI-based python libraries. The protype has been tested by multiple testers against multiple functional and non-functional criteria.

The research has contributed to the existing knowledge by providing a practical and innovative solution to the challenges in the halal food supply chain in ensuring transparency and integrity of halal foods. There are multiple halal food certification and sometimes it might be difficult for the end-consumers to decide. The proposed system provides finer level details about the processes, and activities which makes it easier for consumers to decide the authenticity of the products. As the blockchain records detailed information with the contributions from the three key-stakeholders, the likelihood of mitigating fraud and promoting transparency has also increased. The proposed system will foster consumer confidence and support the growth of the global halal food market.

For future work, this research has only implemented a prototype of the proposed system, the researchers aim to develop a fully functional system. The blockchain infrastructure used in the proposed system has been developed locally by the researchers. For future work, an existing infrastructure or a cloud based blockchain infrastructure (Alrubaiei et al., 2022) might be explored in more details if it provide the level of customization required for the system. Otherwise, a more robust version of the blockchain infrastructure can be developed. The integration of AI and blockchain technology in the proposed system offers several advantages. More AI algorithms can be employed to automate the monitoring and verification processes, thereby enhancing the efficiency and accuracy of halal certification. By leveraging machine learning techniques, the system can also learn from past data and identify patterns or anomalies that may indicate potential non-compliance with halal standards.